# Break-up phenomena of liquid metal thin film induced by high electric current

Rongchao Ma[1], Cangran Guo[2], Yixin Zhou[1], and Jing Liu[1,2]*

[1] *Beijing Key Lab of CryoBiomedical Engineering and Key Lab of Cryogenics, Technical Institute of Physics and Chemistry, Chinese Academy of Sciences, Beijing 100190, China*

[2] *Department of Biomedical Engineering, School of Medicine, Tsinghua University, Beijing 100084, China*

*E-mail: jliu@mail.ipc.ac.cn. Tel.: +86-10-82543765. Fax: +86-10-82543767

**Abstract**

The room temperature liquid metal related electronics has been found important in a wide variety of emerging areas over the past few years. However, its failure features under high electrical current densities are not clear until now. Here we show that a liquid metal thin film would break-up as the applied current increases to a critical magnitude. The break-up phenomenon is attributed to be caused by the so-called electromigration effect. This problem could be one of the major hurdles that must be tackled with caution in the research and application of future liquid metal film electronics.

## 1. Introduction

The room temperature liquid metal has attracted particular attention in recent years due to its fluidity, high electronic conductivity, and large thermal conductivity at normal atmospheric conditions.[1-6] Usually, liquid metal refers to the metals that have low melting points close to room temperature.[7] This property endows the liquid metal a unique feature of easy processing around room temperature. In the light of this merit, the liquid metal has been fund with many important applications such as stretchable electronics[2], printable thermocouple[3], thermometer, cooling media[8], and so on. Recently, it has been gradually extended to direct writing electronics[1,4,5]. In a word, the liquid metal could be a promising material for electronic circuits in the following decades.

However, it is currently not clear as how the reliability of the liquid metal circuits will be when high currents are passing through them. The early studies have shown that an integrated rigid circuit may fail due to the so-called electromigration phenomenon when it carries a high current density.[9-17] This is because the electromigration phenomenon can cause voids and hillocks in the conductor, and therefore, result in an open circuit or a short circuit. Although this problem has been studied in the rigid metal films such as aluminum thin films[18] and copper thin films[19,20] in the field of integrated circuits, little is known about the same issue in the newly emerging liquid metal films. Because such metal has fluidity, the electromigration may cause even more serious problems in the liquid metal circuits. Clarification on the reliability of the liquid metal circuits is therefore critical because it determines the appropriate working of an electronic device thus made. Thus, it is highly desirable to carry out an in-depth investigation on the electromigration phenomenon in the liquid metal thin films.

In this paper, we discovered for the first time the failure phenomenon of liquid gallium thin films. It was experimentally disclosed that a liquid metal thin film breaks up when it carries a high enough current density. The phenomenon was interpreted as a result of electromigration in the liquid metal. This finding is expected to be significant for future studies and applications of liquid metal based printed electronics.

## 2. Experimental set up

The experimental studies on the electromigration phenomenon were carried out in the liquid gallium thin films written on glass and silicon substrates. The liquid gallium used in the experiments has a purity of 99.99%. The glass substrates have dimensions of 76 mm × 25 mm × 1 mm. The samples were prepared by the so-called "direct writing" method[4,5]: cover the substrate by a mask with desired geometric shape, heat the substrate up to 50 - 60 ℃, and then paint the liquid gallium onto the substrate with a brush[3] or glass rod[21]. Because the liquid gallium (and its alloys) need oxides to wet the substrates (glass and silicon)[22,23], the liquid gallium must be repeatedly written at higher temperatures (50 - 60 ℃) to increase its oxide contents. With increasing wettability, the liquid gallium finally sticks to the substrate uniformly.

For the convenience of experimental observation, the thin film samples were prepared with special geometric shapes (see Fig. 1A). The width of the thin film's ends is 7.07 mm and the width of the thin film's middle is 2.32 mm. Consequently, the current density **j** in the middle is more than three times larger than that in the ends. This ensures that the thin films break-up in the middle for the convenient observation under an optical microscope. The thickness of the thin film is 0.012 mm, obtained from the SEM image of the thin film's cross section as shown in Fig. 1B. Due to the insulation of the glass substrate, the sample was coated with an ultrathin layer of gold by sputter deposition for viewing with the SEM.

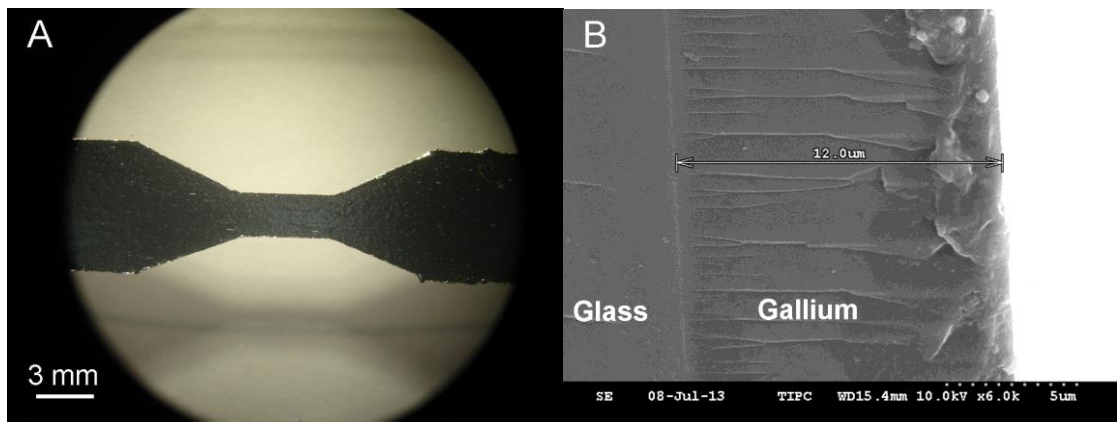

**Figure 1.** A. The geometric shape of the tested liquid gallium thin films. B. The SEM image of the thin film's cross section (the sample is coated with gold).

In the measurements, the electric current was supplied by a current source with a rated current of 10 A. A thermocouple was buried in the liquid metal close to thin film's tapering part to monitor temperature. Because gallium's melting point[7] is $T_m$=29.7646 °C (302.9146 K), we carried out all the measurements at the temperatures above $T_m$ to ensure that our results are from liquid gallium, but not from the rigid solid gallium (see Fig. 3).

### 3. Results

To obtain a visual observation on the high electric current induced break-up phenomenon in liquid gallium thin film, we took the photographs of the thin films under an optical microscope as shown in Fig. 2. Without applying a current to the sample, the middle of the thin film is shown in Fig. 2A (before broken). With an increasing current applied to the sample, the middle of the thin film started to break as shown in Fig. 2B. As the applied current density increased up to j=114.9 A/mm$^2$ (current 3.2 A), the thin film broke up and a crack can be clearly seen on it as indicated in Fig. 2C. The inset of Fig. 2C shows the details of part of the crack. The white region is glass and the black region is liquid gallium. One can also refer to the SEM image of another sample as shown in Fig. 4A to see the detail of the crack. The grayscales of the selected regions in A, B, and C are analyzed as shown D, E, and F, respectively. It indicates that the grayscale becomes light as the thin film started to break. The counts of the dark points is 1100 before the break, but reduced to 650 after the break.

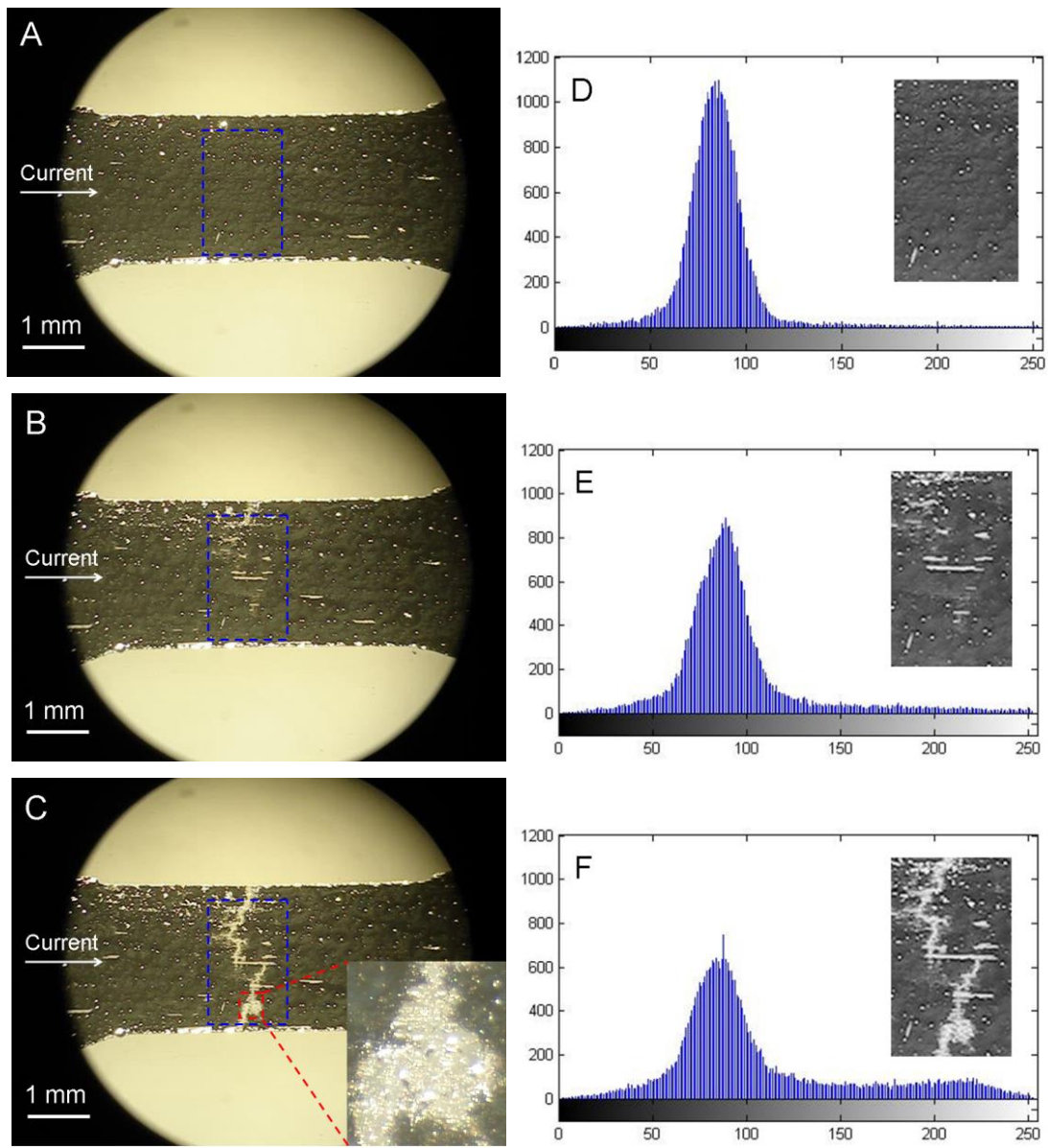

**Figure 2.** The photograph of liquid gallium thin film in the break process caused by electromigration. A. The thin film before break. B. The thin film started to break. C. The thin film after break. D-F are the histograms of the grayscale of the selected regions in A-C, respectively.

The break-up phenomenon as observed in Fig. 2 is further confirmed by monitoring the corresponding electric current density **j**(t), temperature T(t), and resistance R(t) in the thin film as the functions of time t (see Fig. 3):

(1) The electric current density **j**(t) in the thin film is started to increase at t=23.0 s. As **j**(t) increased to 114.9 A/mm$^2$ (current 3.2 A) at t = 26.5 s, it dropped drastically under a constant voltage applied to the sample. Finally, it reached zero at t=27.5 s. This means that the thin film was broken up by the electromigration effect and the circuit became open.

(2) The temperature T(t) in the middle of the thin film is shown to be around 38.0 °C in the interval t < 23.0 s (before break). This ensures that our measurements are performed in liquid gallium but not in solid gallium. With the increasing electric current **j**(t), the temperature then increased to its maximum value T(t)=44.9 °C at t=26.5 s due to the Joule heat released.

(3) The thin film's resistance R(t) between the tapering parts as shown in Fig. 1A increased from 0.4 Ω at t=23.0 s to 0.6 Ω at t=26.5 s, and then jumped up to 97451.1 Ω at t=27.5 s (see the inset of Fig. 3). This indicates that the thin film was broken within one second. Due to the 2D character and the oxide contents, the resistance of the liquid metal thin film is much higher that of the bulk material.

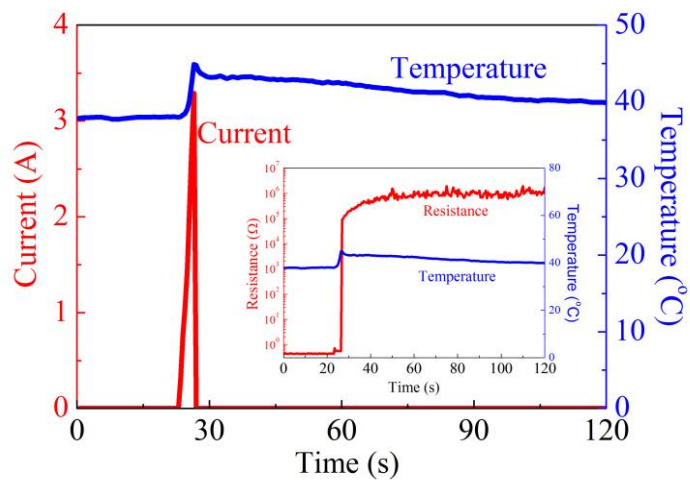

**Figure 3.** Time evolution of the electric current, temperature, and resistance of the liquid metal thin film in the break process. An electronic current was applied to the thin films at t=23.0 s to induce the electromigration. The thin film broke at 27.5 s where the current density increased up to 114.9 A/mm$^2$ (current 3.2 A). The temperature is shown to be around 38 ℃ measured close to the thin film's middle where the break process occurred. This ensured that our measurements are performed in liquid gallium but not in solid gallium. In the break process, the temperature increased from 38 ℃ to 45 ℃ due to the Joule heat released.

The contents of the residue in the crack of the broken thin film were analyzed using EDS (Energy Dispersive Spectroscopy) spectrum technique. To avoid the interference from the oxygen in a glass substrate, we prepared a thin film sample on a silicon substrate. The EDS spectrum was measured by choosing a region in the crack of the broken thin film (see Fig. 4). The elements analysis shows that the remained contains 55.31 Atommic% oxygen, and 44.69 Atomic% gallium. This indicates that the remained are mainly oxides.

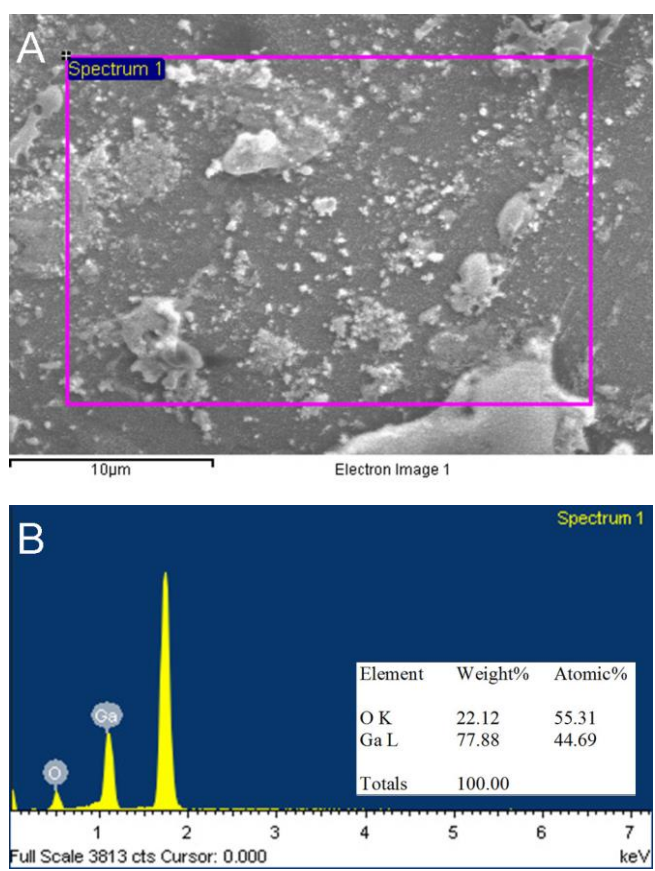

**Figure 4.** EDS spectrum of the liquid gallium thin film on a silicon substrate. The elements analysis indicates that the remained contains 55.31 Atommic% oxygen, and 44.69 Atomic% gallium.

**4. Discussion**

The observed break-up phenomenon in the liquid gallium thin film can be explained using the electromigration theory. It is known that the electromigration phenomenon cause the failure of a solid circuit.[9,10] It is also known that the electromigration phenomenon can induce a liquid metal flow in a micro-channel.[24] On considering that the electromagnetic properties of the liquid metal are similar to those in its solid state, we shall now assume that the electromigration phenomenon will also cause the failure of a conductive liquid circuit.

In the electromigration process,[25] the ion cores are subject to the following two opposing forces as shown in Fig. 5: (1) Electron wind force from the momentum exchange between the conducting electrons and ion cores, $\mathbf{F}_w = Z_w e \mathbf{E}$, where $Z_w$ is the effective valence for the wind force, e is the elementary charge, and $\mathbf{E}$ the electric field.[10,15,16,26-28] (2) Direct electrostatic force from the applied electric field, $\mathbf{F}_d = Z_d e \mathbf{E}$, where $Z_d$ is the effective valence for the electrostatic force.[17] Therefore, the resultant force exerts on the ion cores is the addition of the above two forces, i.e., $\mathbf{F}_{em} = \mathbf{F}_w + \mathbf{F}_d = Z^* e \mathbf{E} = Z^* e \rho \mathbf{j}$, where $Z^* = Z_w + Z_d$ is the effective valence for the resultant force, $\mathbf{j}$ is current density, and $\rho$ the resistivity.[29,30]

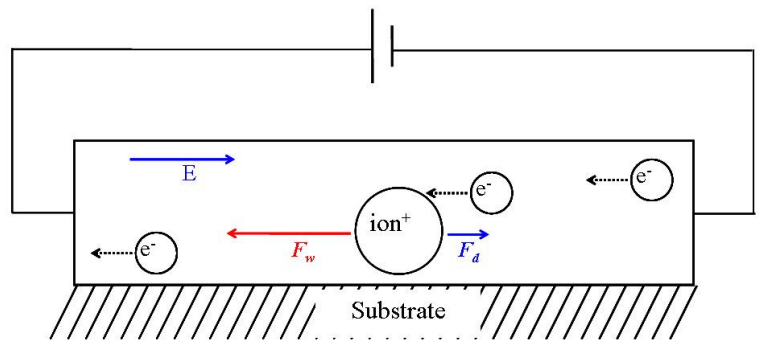

**Figure 5.** The schematic diagram of electromigration in which both the electrons and ion cores move to the opposite direction of the applied electric field.

In addition to the resultant force $F_{em}$, the ion cores in the liquid metal are still subject to other forces.[31-33] All these will contribute to the ion core flow $\mathbf{J}$ in the liquid metal. Thus, the motion of the ion cores in the liquid metal should obey the following modified continuity equation[31-33],

$$\frac{\partial n}{\partial t} + \nabla \cdot \mathbf{J} = 0, \qquad (1)$$

where n is the ion core density and $\mathbf{J}=\mathbf{J}_{em}+\mathbf{J}_n+\mathbf{J}_T+\mathbf{J}_p$ is the ion core mass flux density. The meanings of various terms in **J** are as follows: $\mathbf{J}_{em} = \frac{Dn}{kT}Z^*e\rho\mathbf{j}$ is induced by the electric current density **j**, where $D=\mu kT$ is diffusivity (Einstein relation for liquid), μ is mobility, k is Boltzmann's constant, and T is absolute temperature. $\mathbf{J}_n = -D\nabla n$ is induced by the density gradient $\nabla n$, $\mathbf{J}_T = -\frac{Dn}{kT}\frac{Q}{T}\nabla T$ is induced by the temperature gradient $\nabla T$ (Q is the heat of thermal diffusion), $\mathbf{J}_p = \frac{Dn}{kT}\Omega\nabla p$ is induced by the pressure gradient $\nabla p$ (Ω is the atomic volume).

Eq.(1) shows that, if the flux density moving into a region becomes larger than the flux density moving out from the region $\partial n/\partial t > 0$, then $\nabla \cdot \mathbf{J} < 0$ and material may pile up in this region. However, if the flux density moving into the region becomes smaller than the flux density moving out from the region $\partial n/\partial t < 0$, then $\nabla \cdot \mathbf{J} > 0$ and voids may form in the region. These will cause the circuit failure. Only if $\nabla \cdot \mathbf{J} = 0$ (or $\partial n/\partial t = 0$), then there is no failure occurs. This can well explain the results in Fig. 2 and Fig. 3.

There is another possibility that the break-up phenomenon in the liquid metal thin film may be caused by a temperature distribution. However, our early measurements have shown that the thin film stays unaffected even as the temperature in the thin film's middle increased up to 375 °C. This ruled out the possibility of thermal break-up in our liquid metal thin film. In the following discussion, therefore, we shall focus on the effects of the electromigration, but ignore the others, i.e., $\mathbf{J}_{em} \neq 0$, $\mathbf{J}_T \approx \mathbf{J}_p \approx \mathbf{J}_n \approx 0$.

Another measuring procedure is also used to confirm that the break-up phenomenon in the liquid metal thin film is induced by the electric current **j** (or electron wind force $\mathbf{F}_w$), but not by the applied electric field **E** (or direct electrostatic force $\mathbf{F}_d$). More specifically, apply an electric field **E** (or a voltage U) to the sample, but let the circuit open to ensure that no electric current is observed (**j**=0) in the circuit. Our results have shown that no break-up phenomenon was observed even if the voltage was increased to a higher value. This can be well understood by referring to Eq.(1). If **j**=0, then $\nabla \cdot \mathbf{J}_e = 0$. Consequently, we have $\partial n/\partial t = 0$. Therefore, the ion core density is a constant and no electromigration phenomenon will occur.

The EDS spectrum results in Fig. 4 may suggest that the electron wind force $\mathbf{F}_w$ is small in the few layers of atoms absorbed on the substrate.[24] It is known that gallium and its alloys need oxides to wet the substrates such as glass and silicon[22,23], the thin layer between the liquid metal and substrate are mainly oxides ($Ga_2O_3$) that are wide bandgap semiconductors. This layer should not be subject to the electron wind force $\mathbf{F}_w$ because there is no transporting electron within the oxides. In the electromigration process, therefore, the oxides should stay behind because there is no electron flow inside it, but the liquid gallium on the top were pushed away by the electron flow while the oxides underneath stay behind.

Due to its fluidity, therefore, a liquid metal thin film is more vulnerable to the electromigration phenomenon than a solid thin film. Thus, the electromigration phenomenon should have a considerable impact on the reliability of the liquid metal film in printed electronics, electronic circuit, 3-D electronic fabrication or related microtechnology etc., which must be handled with caution.

Finally, we would like to mention that, in addition to Eq.(1), the liquid gallium should obey two more modified equations of fluid mechanics:

*Equation of motion (Navier–Stokes equation)*[34,35]. Referring to the general form of the equation of motion (Cauchy momentum equation) and considering the resultant force $\mathbf{F}_{em}=Z^*e\rho\mathbf{j}$, one can easily obtain the Navier–Stokes equation for a liquid metal that includes the electromigration effect,

$$n\left[\frac{\partial \mathbf{v}}{\partial t}+(\mathbf{v}\cdot\nabla)\mathbf{v}\right]=-\nabla p+\eta\Delta\mathbf{v}+\left(\frac{1}{3}\eta+\zeta\right)\nabla(\nabla\cdot\mathbf{v})+nZ^*e\rho\mathbf{j}, \qquad (2)$$

where, $\mathbf{v}$ is fluid velocity, p the pressure in the liquid metal, $\eta$ the first coefficient of viscosity, and $\zeta$ the second coefficient of viscosity (or volume viscosity).

*Energy equation*[34,35]. By considering the conservation of energy, one can obtain the following equation,

$$\frac{d}{dt}\left(n\varepsilon+\frac{1}{2}n\mathbf{v}^2\right)=\nabla\cdot[(-p\mathbf{I}+\boldsymbol{\tau})\cdot\mathbf{v}]+\nabla\cdot(\kappa\nabla T)+\rho j^2, \qquad (3)$$

where, $\varepsilon$ is the internal energy per unit mass, $\mathbf{I}$ is the identity tensor, $\boldsymbol{\tau}$ the viscous stress, i.e., $\tau_{ij}=\eta\left(\frac{\partial v_i}{\partial x_j}+\frac{\partial v_j}{\partial x_i}-\frac{2}{3}\delta_{ij}\nabla\cdot\mathbf{v}\right)+\varsigma\delta_{ij}\nabla\cdot\mathbf{v}$. The term $\nabla\cdot(\kappa\nabla T)$ represents the heat flowed out from the region and $\kappa$ is the thermal conductivity. The term $\rho j^2 = \mathbf{E} \cdot \mathbf{j}$ is the Joule heat released, which includes the energy increased in the electromigration process.

Furthermore, the liquid metal must also obey Maxwell's equations, which are not given here for brief.

A complete characterization on the above equations is beyond the scope of this study. Overall, the present findings also raised quite a few important fundamental issues worth of pursuing in the coming time.

## 5. Conclusions

In summary, we have discovered for the first time that liquid metal is susceptible to electromigration effect. A liquid metal thin film can be broken up by electron wind force when it carries a high current density. These findings open a new direction for the study of the electromigration phenomenon in conductive fluids which had never been reported before[36]. The electromigration phenomenon also adds a barrier for the large scale applications of the liquid metal in electronics. Further studies need to be carried out to resolve the failure problem in liquid metal film caused by the electromigration phenomenon.